\documentclass[11pt,twoside]{article}

\usepackage{asp2010}
\usepackage[T1]{fontenc}
\usepackage[latin1]{inputenc}
\usepackage[ngerman,english]{babel}

\resetcounters

\begin{document}

\title{Kepler light curves and stellar rotational periods}
\author{Timo Reinhold$^1$, Ansgar Reiners$^1$, Gibor Basri$^2$, and Lucianne M. Walkowicz$^2$}
\affil{$^1$Universität Göttingen, Institut für Astrophysik, Friedrich-Hund-Platz 1, 37077 Göttingen, Germany}
\affil{$^2$Astronomy Department, University of California, Berkeley, CA 94720, USA}

\begin{abstract}
The Kepler space telescope monitors over 156.000 stars with an unprecedented photometric precision. We are interested in stellar rotational periods which we find using Lomb-Scargle periodograms. This work focuses on the 306 exoplanet candidate host stars released on June 15, 2010. We present statistics on how many of them show periodic photometric variability, providing preliminary periods and estimates of stellar activity. In the future, our work will focus on spot evolution and differential rotation.
\end{abstract}

\section{Introduction}
Periodic stellar variability has three main sources: Binarity (gravitational deformation, eclipses), surface oscillations, and rotation (due to star spots). The variability induced by each process has a characteristic range of timescales and amplitudes. In this work, we are interested in studying rotation-induced variability on Kepler stars. The signal we are planning to detect should have a periodicity in the range of 0.5--17 days and a relative flux drop of the order of 0.5--15 \%, depending on how fast the stars are rotating and on the relative size of the spots on their surfaces. To find these periodicities we are using the Lomb-Scargle periodogram. Our main goal is to derive stellar rotational periods. Since our algorithm gives no information about the source of the flux variations all derived periods should be taken as preliminary results.

\section{Kepler light curves}
\begin{figure}
	\begin{center}
	  $\textrm{P}=11.84\pm1.66$ d
	  \includegraphics[angle=90, scale=0.437]{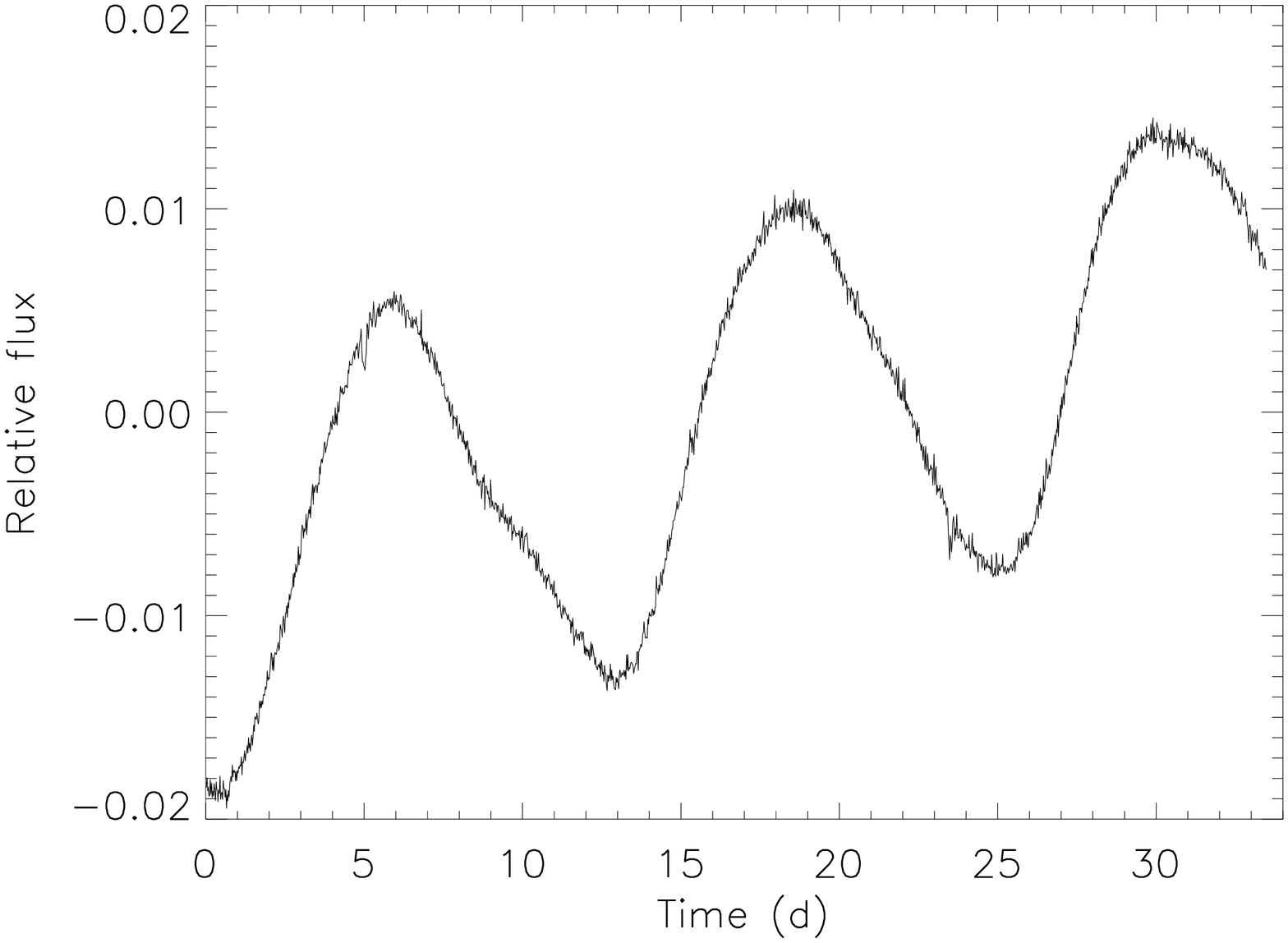}
	\end{center}
	\caption{Light curve of the star with Kepler-ID: 5812960, $T_{\rm eff}$=5117 K, $\log$ g=4.408 and KEP-Mag: 14.92}
	\label{lc_136}
\end{figure}

\begin{figure}
	\begin{center}
	  $\textrm{P}=11.19\pm1.57$ d
	  \includegraphics[angle=90, scale=0.437]{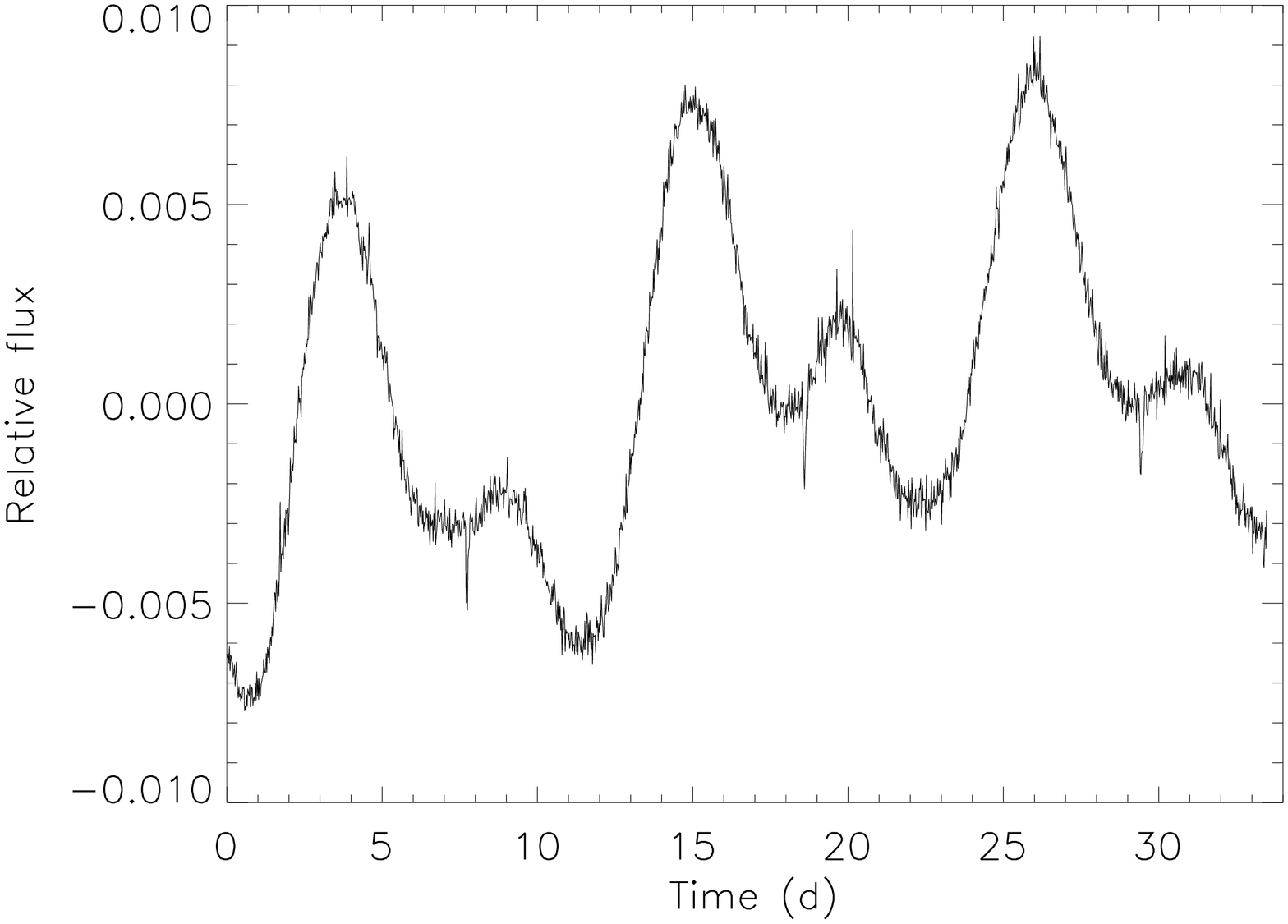}
	\end{center}
	\caption{Light curve of the star with Kepler-ID: 8505670, $T_{\rm eff}$=4214 K, $\log$ g=4.608 and KEP-Mag: 15.06}
	\label{lc_302}
\end{figure}

\begin{figure}[!ht]
	\begin{center}
	  $\textrm{P}=1.02\pm0.02$ d
	  \includegraphics[angle=90, scale=0.437]{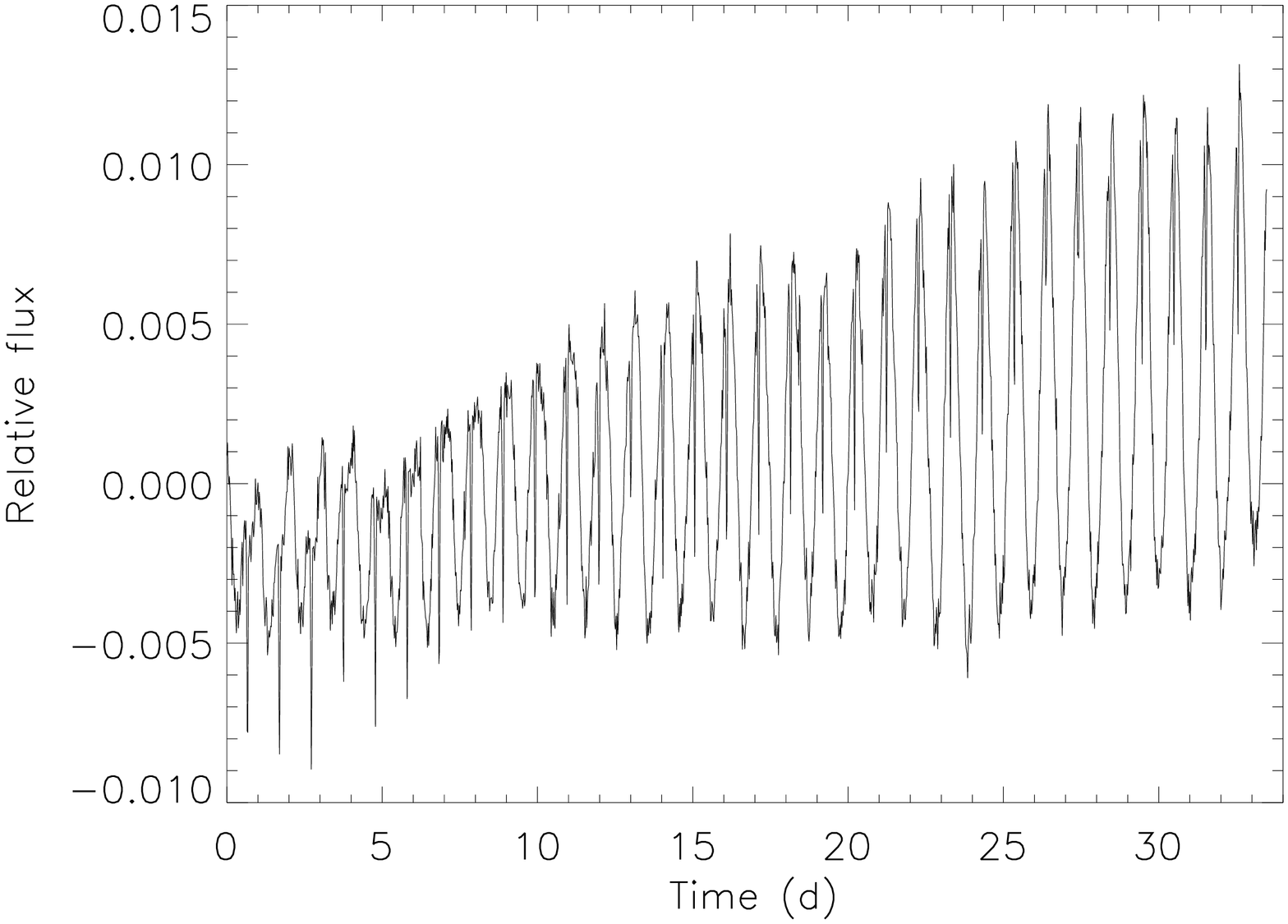}
	\end{center}
	\caption{Light curve of the star with Kepler-ID: 5115978, $T_{\rm eff}$=5976 K, $\log$ g=4.427, KEP-Mag: 15.20 and the shortest detected period, probably due to pulsations.}
	\label{lc_46}
\end{figure}

\begin{figure}
	\begin{center}
	  $\textrm{P}=8.91\pm1.01$ d
	  \includegraphics[angle=90, scale=0.437]{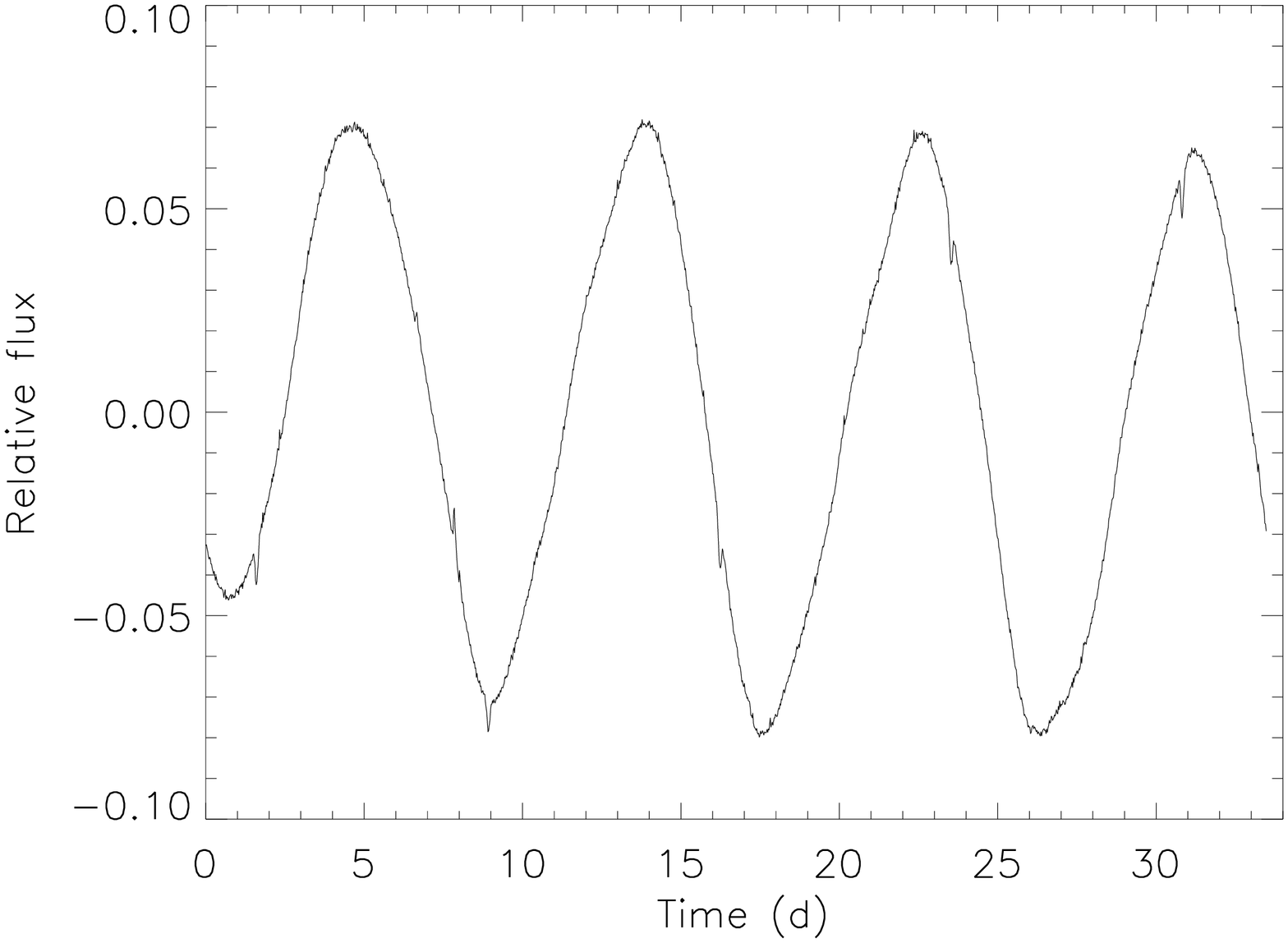}
	\end{center}
	\caption{Light curve of the star with Kepler-ID: 10068383, $T_{\rm eff}$=5046 K, $\log$ g= 4.652 and KEP-Mag: 15.77}
	\label{lc_269}
\end{figure}

Figs.\ref{lc_136}-\ref{lc_269} present a variety of beautiful Kepler light curves all showing by eye clear periodicity. We divided each light curve by its median and subtracted unity. In almost all light curves a linear trend is seen likely caused by the instrument (in Fig.\ref{lc_269} this trend was removed). All calculations we did are based on the so-called raw data (marked as ``AP\_RAW\_FLUX'' in the FITS files) except for Fig.\ref{lc_269} where we used the linear flattened light curve. The values of the unique Kepler-ID, $T_{\rm eff}, \ \log g$ and apparent magnitude are taken from the Kepler Input Catalogue (KIC) and the light curves can be found at http://archive.stsci.edu/pub/kepler/lightcurves/tarfiles/.

\section{Lomb-Scargle periodogram}
The Lomb-Scargle periodogram \citep{scargle} is a powerful spectral analysis tool for unevenly sampled data. It decomposes the data into a series of sines and cosines and one obtains peaks with different powers depending on the goodness of the fit. The periods we found are always associated to the highest peak of the periodogram. Futhermore, we overplotted the highest peak with a Gaussian and took the half of the full width at half maximum to determine the period error. This is a very conservative approach and the errors seem to be quite large. If one takes the time span and the number of data points into account the period errors can be decreased dramatically (see \citet{Gilliland1985} or \citet{Montgomery1999}). \\
To test the reliability of the periods returned by the algorithm we introduced noise to our light curves. Furthermore, we checked the influence of the transits contained in our sample on the period determination. It turns out that the Lomb-Scargle periodogram is very stable under these flux perturbations and that transits don't cause any effect because the number of their data points is very small.

\section{Sample}
\begin{figure}[!ht]
	\begin{center}
	  \includegraphics[angle=90, scale=0.437]{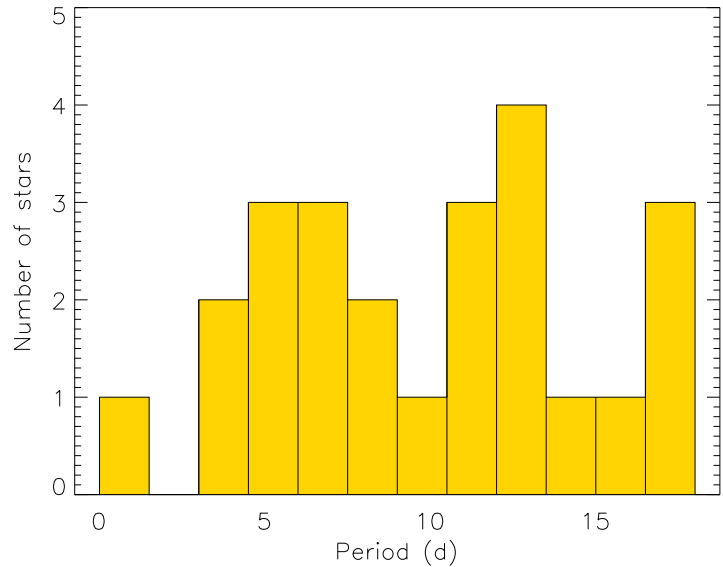}
	\end{center}
	\caption{Period distribution of the 24 active stars}
	\label{histo_70}
\end{figure}

The Kepler Quarter 1 (Q1) data was released to the public on June 15, 2010. It has a time span of $\sim$ 33.5 days and a cadence of $\sim$ 30 minutes. 
Our sample consists of the 306 exoplanet candidate host stars having transit events in their light curves. Effective temperature of the sample ranges from $4000-6300$ K and $\log$ g is between $4.2$ and $5.1$. A statistical overview of the whole Kepler sample can be found in \citet{basri2,basri1}. \\
Since we are interested in stellar rotational periods, we selected a subsample of 24 ``active'' stars from the whole data set that have periodogram power greater than 70 and periods up to 17 days. These restrictions account for the facts that the light curve should show by eye a believable periodicity and that at least two full cycles are observed. The period distribution of our subsample is shown in Fig.\ref{histo_70}. Most of the periods are believed to be stellar rotational periods but in some cases it is possible that just the half period was detected.

\section{Summary \& Future work}
Using the Lomb-Scargle periodogram, we selected some active stars from our sample and determined preliminary periods and their errors. Assuming the flux variations are due to spots, one has to be very careful to assign these periods as true stellar rotational periods since the spot distribution is unknown. Furthermore, star spots use to change their size and position, some are created, others disappear during a cycle. To account for all possible situations, explicit spot modeling needs to be done. \\
The light curve in Fig. \ref{lc_46} shows some kind of beat frequency. The duration of the signal indicates that it is probably due to pulsations. But also stars exhibiting differential rotation should show similar light curves. Therefore, our future work will focus on the evolution of periods over time, especially considering differential rotation. As more data becomes available we will also look for activity cycles.

\acknowledgements T. R. is supported by a scholarship of the DFG Graduiertenkolleg 1351 ``Extrasolar Planets and their Host Stars''. A.R. acknowledges support from the DFG under RE 1664/4-1.
\bibliography{reinhold_t}

\newpage

\begin{figure}
	\begin{center}
	  $\textrm{P}=8.91$ d, $\textrm{Power}=195.1$
	  \includegraphics[angle=90, scale=0.437]{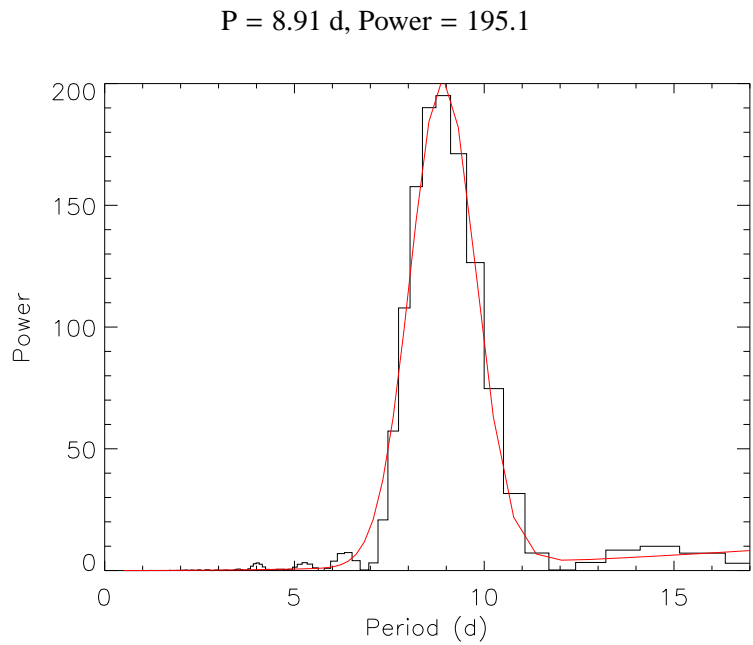}
	\end{center}
	\caption{Periodogram of light curve from Fig. \ref{lc_269}}
	\label{per_269}
\end{figure}

\end{document}